# Interaction of a three-level atom (Λ,*V*, Ξ) with a two-mode field beyond rotating wave approximation: Intermixed intensity-dependent coupling

N. Asili Firouzabadi, M. K. Tavassoly

Optics and Laser Group, Faculty of Physics, University of Yazd, Yazd, Iran

ABSTRACT—Recalling that the rotating wave approximation (RWA) is only valid in the weak coupling regimes, the purpose of this paper is to study the Hamiltonian dynamics describing the full quantum mechanical approach of the interaction between various configurations of three-level atoms Λ,*V* and Ξ distinctly with a two-mode radiation field, while the RWA is not considered; i.e., the counter-rotating terms (CRTs) are taken into account. Generally, the presence of CRTs in the Hamiltonian prevents one to achieve an analytical solution. Moreover, as we will show in the present work, using the perturbation theory, analytical solvable Hamiltonians can be successfully obtained. According to our calculations, the contribution of CRTs within the ordinary Hamiltonian is equivalent to the replacement of the 'constant detuning' with a specific 'intensity-dependent detuning' in the first order, and the 'constant atom–field coupling' with a particular intensity-dependent ($f$-deformed) coupling in the second order of the associated perturbation parameter. Moreover, noticing that according to the initial expression of the Hamiltonian, each mode of the field interacts only with a specific pair of the allowed transitions of each type of the three-level atom, it is surprisingly observed that via applying the mentioned approach, the obtained intensity-dependent coupling functions depend on both modes of the field, i.e., $f(\hat{n}_a,\hat{n}_b)$. In this way, it is seen that the CRTs are removed with the price of arriving at some intermixed intensity-dependent atom-field coupling functions of the two modes of the field. In this way, the obtained final Hamiltonians are analytically solvable. At last, by determining the time evolution of the atom–field wave function, we study the effects of CRTs on a few nonclassical properties of the state of the system, including the atomic population inversion and photon statistics.

I. Introduction

For the first time, in 1963, the Jaynes–Cummings model (JCM) propounds a complete quantum approach for the description of the interaction of a two-level atom with a single-mode electromagnetic field [1]. Afterwards, to study on more complex atom-field interacting systems, various generalizations of JCM have been introduced in the literature, including the extension to three-level systems [2] (like Λ, V, Ξ type) interacting with quantized fields [3], various atom-field interacting systems with intensity-dependent couplings [4] and so on. We should emphasize that in JCM, rotating wave approximation (RWA) is taken into account; the fact that enables one to obtain an analytic solution for the Hamiltonian dynamics, while without it, usually one cannot solve the problem. To explain more, the RWA is a very popular technique in quantum optics, laser physics and other branches of resonance phenomena, since it leads to some mathematical simplifications in the calculation procedures of many problems. In this line, it is well-known that, the RWA is legitimized to small detuning and small ratios of the atom-field coupling divided by the atomic transition frequency [5, 6, 7]. Altogether, over the recent decades, new coupling regimes of the quantum Rabi model have been investigated, in which the coupling strengths of interaction are greater than the frequencies of the components without interaction [8, 9]. Such phenomena are typically observed experimentally in several laboratory systems [10, 11]. In this regard, experimental progresses in the field reveal that for some of systems like trapped ions [12], cooper-pair boxes [13, 14] and flux qubits [15] the coupling can be very large (i.e., we have encountered strong-coupling regimes), and so the RWA clearly breaks down [16].

Based on the above-mentioned observations, CRTs (virtual-photon processes) receive increasing attentions of the authors in various systems, since they can yield substantial and interesting physical effects. Among them, cavity systems with strong couplings have been studied [17]. Peng and Li have numerically analyzed the effect of the virtual-photon field on phase fluctuation and the periodic collapse revivals of atomic population in the JCM [18, 19, 20]. In [21] the entanglement dynamics of two atoms interacting with a dissipative coherent cavity field without RWA has been investigated. The entanglement evolution of two independent atoms without RWA has been proposed in [22] and the influence of CRTs in the JCM on the atomic population inversion has been investigated in [23]. It is shown that, the inclusion of CRTs to the Hamiltonian results in a transient entanglement between two atoms coupled to a standing-wave single-mode cavity field and it is found that the main source of entanglement is the two-photon coherence induced by two-photon transitions through virtual states [24].

Since the RWA cannot be well applicable in the strong-coupling regimes, therefore, the solution procedure should proceed in the presence of CRTs which cause serious problems in the analytical solution of the system. Moreover, CRTs may be smoothly removed by perturbation theory approach [25, 26]. In this regard, recently in [27, 28], an analytical method, based on perturbation theory, has been presented to describe the interaction between a two-level atom and a single-mode field in the absence of RWA. It is shown that, the constant atom-field coupling and detuning parameters in the atom-field interaction without RWA have been converted to intensity-dependent terms, but fortunately with RWA. Compared with the two-level atoms with a unique possible transition, three-level atoms ($\Lambda, V, \Xi$) have two different transitions between their energy levels. Therefore, more diverse quantum phenomena may be expected from the three-level atoms [29]. Therefore, it seems that it is necessary to deeply study the three-level atoms either in weak or strong atom-field coupling. More recently, using the same approach in [27, 28], in Ref. [30] the authors studied the interaction of a $\Lambda-$type three-level atom, however still with a single-mode radiation field without RWA. However, keeping in mind that in this case the atom possesses three levels with two allowed transitions between the states, there are two different coupling constants, namely $g_1$. $g_2$. As a result, considering CRTs, these two coupling constants have been changed respectively to two specific intensity-dependent coupling functions $f_1(\hat{n}), f_2(\hat{n})$ which clearly depend on the number of photons. It is also observed that the nonlinearity function $f_1(\hat{n})$ ($f_2(\hat{n})$) not only depends on $g_1(g_2)$ but also on $g_2(g_1)$.

In recent decades, the interaction between atom and field is widely studied.

In this line, in particular, to be right, the interaction between a three-level atom with a single-mode field (in the presence of detunings) is not physically well understandable. Therefore, in this paper we aim to develop the work in [30] to two-mode field. In more detail, due to the importance of CRTs in the strong coupling regimes, in this study, according to our previous explanations, we encouraged to investigate the interaction of three-level atoms ($\Lambda, V, \Xi$ type) individually with a two-mode field without RWA. Our calculations yield three different intensity-dependent functions for the three configurations of three-level atoms. Briefly speaking, the two coupling constants $g_1, g_2$ have been changed respectively to two distinct intensity-dependent coupling functions $f_1(\hat{n}_1, \hat{n}_2)$, $f_2(\hat{n}_1, \hat{n}_2)$ which clearly depend on the number of photons of the two modes. It may be observed that the nonlinearity function $f_1(\hat{n}_1, \hat{n}_2)$ ($f_2(\hat{n}_1, \hat{n}_2)$) not only depends on $\hat{n}_1$, $g_1(n_2, g_2)$ but also depends on $\hat{n}_2, g_2$ ($\hat{n}_1, g_1$). In this way, we arrive at two intertwined nonlinearity functions for each type of three-level atoms, when they interact with a two-mode field. At this point it should be mentioned that atom-field intensity dependent function has a long background in the literature started by Buck and Sukumar [31,32] and followed by others [33-35].

We end this section with mentioning that, our used approach simply transforms the unsolvable Hamiltonians to some solvable ones. Therefore, we can easily investigate the effect of CRTs on the physical properties of the obtained state vector of the system. To do this task, we pay our attention to the atomic population inversion and the photon statistics of the field, which clarify the nonclassicality feature of the quantum state.

The paper is organized as follows. In section 2 we solve analytically the dynamics of the interaction between all types of three-level atoms with a two-mode field without RWA, using the perturbation theory. In the next section, we discuss the effect of CRTs on the dynamics of the atomic population inversion and the Mandel parameter. Finally, in section 4 we present a summary and concluding remarks.

## II. THE INTERACTIONS OF THREE-LEVEL ATOMS WITH TWO-MODE FIELDS AND THEIR ANALYTICAL SOLUTIONS

The purpose of this paper is to solve the dynamics of various configurations of three-level atoms ($\Lambda, V, \Xi$) which distinctly interact with a two-mode quantized field, while the RWA is not considered (or equivalently, CRTs are taken into account). We suppose that the atom-field coupling is so strength that one cannot ignore the CRTs. As a result, at first glance it seems that the associated interaction Hamiltonians cannot be solved analytically. However, in the continuation, via following the perturbation theory approach, analytical solution for the Hamiltonian dynamics of each type of the atoms can be achieved.

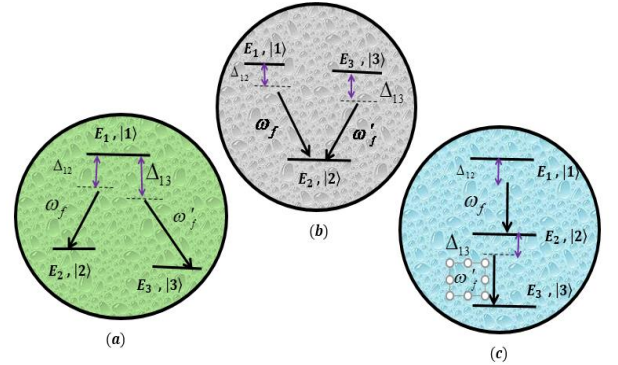

Fig. **1** Three-level atomic structure for (a) $\Lambda$-type, (b) $V$-type and (c) $\Xi$ type, with two-mode field.

### 2 · 1 $\Lambda$ − *type three-level atom without RWA*

We begin with the $\Lambda$-type three-level atom, where its atomic levels are labeled as the ground states $|2\rangle$, $|3\rangle$ and the excited state $|1\rangle$. The transitions between ground states $|2\rangle$ and state $|3\rangle$ are forbidden, while between $|1\rangle \leftrightarrow |3\rangle$ and $|1\rangle \leftrightarrow |2\rangle$ are allowed in the dipole approximation (see figure 1 (a)). In fact, we introduce a model in which a two-mode quantized radiation field interacts with a $\Lambda$-type atom. The Hamiltonian of the system (containing the CRTs) with electric dipole approximation reads as ($\hbar = 1$):

$$\hat{H} = \hat{H}_0 + \hat{H}_1 + \hat{H}_2, \tag{1}$$

where

$$\hat{H}_0 = \omega_f \hat{a}^\dagger \hat{a} + \omega_{f'} \hat{b}^\dagger \hat{b} + \sum_{j=1,2,3} E_j \hat{\sigma}_{jj}, \tag{2}$$

$$\hat{H}_1 = g_1(\hat{a}^\dagger \hat{\sigma}_{21} + \hat{a} \hat{\sigma}_{12}) + g_2(\hat{b}^\dagger \hat{\sigma}_{31} + \hat{b} \hat{\sigma}_{13}), \tag{3}$$

$$\widehat{H}_2 = g_1(\hat{a}^\dagger \hat{\sigma}_{12} + \hat{a}\hat{\sigma}_{21}) + g_2(\hat{b}^\dagger \hat{\sigma}_{13} + \hat{b}\hat{\sigma}_{31}), \tag{4}$$

where $E_j$ is the energy of jth atomic level, $\hat{\sigma}_{ij} = ij(i,j=1,2,3)$ denotes the atomic ladder operator between levels $i,j$, $\hat{a}$ and $\hat{a}^\dagger$ ($\hat{b}$, $\hat{b}^\dagger$) are respectively the bosonic annihilation and creation operators of the field, $\omega_f$ and $\omega_{f'}$ are the frequencies of the field modes and $g_1$, $g_2$ denote the coupling strengths between the field and the two allowed transitions.

Usually in the literature, $\widehat{H}_2$ which consists of the CRTs is neglected; this is known as the RWA. There exist some reasons for the interests in this approximation, among them, lacking an analytical solution of the Hamiltonians containing the CRTs is remarkable. Moreover, in large coupling strengths, the CRTs are not negligible. Therefore, our purpose of the paper is to arrive an analytical solution to the problem while the CRTs are also maintained. To achieve the purpose, inspired from Refs. [27, 28, 29] we apply the following three unitary transformations on $\widehat{H}$ in (1), successively. At first,

$$\widehat{U}_1 = e^{\varepsilon_1(\hat{a}^\dagger\hat{\sigma}_{12} - \hat{a}\hat{\sigma}_{21}) + \varepsilon_2(\hat{b}^\dagger\hat{\sigma}_{13} - \hat{b}\hat{\sigma}_{31})}, \tag{5}$$

where

$$\varepsilon_1 = \frac{g_1}{2\widetilde{\omega}_{12}}, \widetilde{\omega}_{12} = \frac{E_1 - E_2 + \omega_f}{2},$$

$$\varepsilon_2 = \frac{g_2}{2\widetilde{\omega}_{13}}, \quad \widetilde{\omega}_{13} = \frac{E_1 - E_3 + \omega_{f'}}{2}, \tag{6}$$

and then,

$$\widehat{U}_2 = e^{\beta_1(\hat{a}^{\dagger 2} - \hat{a}^2)\hat{\sigma}_z^{12} + \beta_2(\hat{b}^{\dagger 2} - \hat{b}^2)\hat{\sigma}_z^{13}}, \tag{7}$$

where

$$\beta_1 = \frac{\varepsilon_1 g_1}{2\omega_f}, \beta_2 = \frac{\varepsilon_2 g_2}{2\omega_{f'}}, \tag{8}$$

and finally,

$$\widehat{U}_3 = e^{\gamma_1(\hat{a}^{\dagger 3}\hat{\sigma}_{21} - \hat{a}^3\hat{\sigma}_{12}) + \gamma_2(\hat{b}^{\dagger 3}\hat{\sigma}_{31} - \hat{b}^3\hat{\sigma}_{13})}, \tag{9}$$

where

$$\gamma_1 = \frac{2\varepsilon_1^2 \widetilde{\omega}_{12} + \varepsilon_2^2 \widetilde{\omega}_{13}}{\omega_f(E_1 - E_2 - 3\omega_f)} g_1,$$

$$\gamma_2 = \frac{2\varepsilon_2^2 \widetilde{\omega}_{13} + \varepsilon_1^2 \widetilde{\omega}_{12}}{\omega_{f'}(E_1 - E_3 - 3\omega_{f'})} g_2, \tag{10}$$

With the help of the preceding transformations and through a lengthy calculations, we arrive at the following effective Hamiltonian ($\hat{n}_1 = \hat{a}^\dagger\hat{a}$ and $\hat{n}_2 = \hat{b}^\dagger\hat{b}$ are the number operators):

$$\widehat{H}_{eff} = \widehat{U}_3^\dagger \widehat{U}_2^\dagger \widehat{U}_1^\dagger \widehat{H} \widehat{U}_1 \widehat{U}_2 \widehat{U}_3 \approx \widehat{H}_0 + \widehat{H}_1^p$$

$$+\varepsilon_1 g_1 [(\hat{n}_1 + \frac{1}{2})\hat{\sigma}_z^{12} + \frac{\hat{\sigma}_{33}}{2} - \frac{1}{2}]$$

$$+\varepsilon_2 g_2 \left[\left(\hat{n}_2 + \frac{1}{2}\right)\hat{\sigma}_z^{13} + \frac{\hat{\sigma}_{22}}{2} - \frac{1}{2}\right], \tag{11}$$

where $\widehat{H}_1^D$ in (11) is introduced as,

$$\widehat{H}_1^D = \hat{a} g_1 f_1(\hat{n}_1,\hat{n}_2)\hat{\sigma}_{12} + \hat{a}^\dagger\hat{\sigma}_{21} \tag{12}$$

$$+\hat{b} g_2 f_2(\hat{n}_1,\hat{n}_2)\hat{\sigma}_{13} + g_2 f_2(\hat{n}_1,\hat{n}_2)\hat{b}^\dagger\hat{\sigma}_{31}, \tag{12}$$

with the following definitions of the nonlinearity functions corresponding to intensity-dependent atom-field couplings,

$$f_1(\hat{n}_1,\hat{n}_2) = \left(1 - \varepsilon_1^2 \hat{n}_1 - \frac{\varepsilon_2^2}{2}\hat{n}_2\right),$$

$$f_2(\hat{n}_1,\hat{n}_2) = \left(1 - \varepsilon_2^2 \hat{n}_2 - \frac{\varepsilon_1^2}{2}\hat{n}_1\right). \tag{13}$$

It may be emphasized that in our above calculations, higher orders of the perturbation parameters ($\varepsilon_1, \varepsilon_2 \ll 1$) have been neglected according to the perturbation approach. Also, in applying the third transformation, it is assumed that we are far from three-photon transition; this means that $E_1 - E_2 = 3\omega_f, E_1 - E_3 = 3\omega_{f'}$. In this way, we arrived at an effective Hamiltonian that preserves the total excitation of quanta. This is the main advantage of the approach. By this we mean that, even though we encountered the CRTs in the Hamiltonian (see the initial Hamiltonian in Eqs. (1)-(4), moreover, an analytical solution can be reasonably expectable, just like the case in which we encountered the RWA in the atom-field systems. Also note that, in the effective Hamiltonian (11), the CRTs have been vanished and in the first-order Hamiltonian (of $\varepsilon_1$ and $\varepsilon_2$), the third and fourth terms appear which describes the quantum shift in the energy of the atom–field system. In addition, it is noticeable that, in the second-order Hamiltonian (of $\varepsilon_1$ and $\varepsilon_2$), the considered coupling constants $g_1$ and $g_2$ have been converted to $g_1 f_1(\hat{n}_1,\hat{n}_2)$ and $g_2 f_2(\hat{n}_1,\hat{n}_2)$, respectively. Therefore, briefly, we showed that considering CRTs (i.e., JCM beyond RWA) our used approach straightforwardly transforms the unsolvable linear (constant atom-field coupling) interaction dynamics to a solvable nonlinear (intensity-dependent) coupling JCM.

## 2.2 $V-$ and $\Xi$ type three-level atoms with two-mode field without RWA

After presenting the detailed procedure of the interaction between a $\Lambda$-type atom with a two-mode field, containing the CRTs, in the previous subsection, in this subsection we want to consider the other two types of atoms, shortly. At first, we pay our attention to the $V$-type three-level atom where its two upper levels is allowed to transit to the lower level, and the transition between the two upper levels is forbidden (see figure 1(b)). By this description, the full Hamiltonian of the interacting system (containing the CRTs) in the electric dipole approximation reads as ($\hbar=1$):

$$\widehat{H} = \widehat{H}_0 + \widehat{H}_1 + \widehat{H}_2, \qquad (14)$$

where $\widehat{H}_0$ has been introduced in Eq. (2) and

$$\widehat{H}_1 = g_1(\hat{a}^\dagger \hat{\sigma}_{21} + \hat{a}\hat{\sigma}_{12}) + g_2(\hat{b}^\dagger \hat{\sigma}_{23} + \hat{b}\hat{\sigma}_{32}), \qquad (15)$$

$$\widehat{H}_2 = g_1(\hat{a}^\dagger \hat{\sigma}_{12} + \hat{a}\hat{\sigma}_{21}) + g_2(\hat{b}^\dagger \hat{\sigma}_{32} + \hat{b}\hat{\sigma}_{23}), \qquad (16)$$

As in the $\Lambda$-type atom, however via applying the following three successive unitary transformations,

$$\widehat{U}_1 = e^{\varepsilon_1(\hat{a}^\dagger \hat{\sigma}_{12} - \hat{a}\hat{\sigma}_{21}) + \varepsilon_2(\hat{b}^\dagger \hat{\sigma}_{32} - \hat{b}\hat{\sigma}_{23})}, \qquad (17)$$

$$\widehat{U}_2 = e^{\beta_1(\hat{a}^{\dagger 2} - \hat{a}^2)\hat{\sigma}_z^{12} + \beta_2(\hat{b}^{\dagger 2} - \hat{b}^2)\hat{\sigma}_z^{23}}, \qquad (18)$$

$$\widehat{U}_3 = e^{\gamma_1(\hat{a}^{\dagger 3}\hat{\sigma}_{21} - \hat{a}^3 \hat{\sigma}_{12}) + \gamma_2(\hat{b}^{\dagger 3}\hat{\sigma}_{23} - \hat{b}^3 \hat{\sigma}_{32})}, (19)$$

on the Hamiltonian (14), the following effective Hamiltonian may be obtained,

$$\widehat{H}_{eff} \approx \widehat{H}_0 + \widehat{H}_1^D$$
$$+\varepsilon_1 g_1 [(\hat{n}_1 + \tfrac{1}{2})\hat{\sigma}_z^{12} + \tfrac{\hat{\sigma}_{33}}{2} - \tfrac{1}{2}]$$
$$+\varepsilon_2 g_2 \left[(\hat{n}_2 + \tfrac{1}{2})\hat{\sigma}_z^{32} + \tfrac{\hat{\sigma}_{22}}{2} - \tfrac{1}{2}\right], \qquad (20)$$

with

$$\widehat{H}_1^D = \hat{a} g_1 f_1(\hat{n}_1,\hat{n}_2)\hat{\sigma}_{12} + g_1 f_1(\hat{n}_1,\hat{n}_2)\hat{a}^\dagger \hat{\sigma}_{21}$$
$$+\hat{b} g_2 f_2(\hat{n}_1,\hat{n}_2)\hat{\sigma}_{32} + g_2 f_2(\hat{n}_1,\hat{n}_2)\hat{b}^\dagger \hat{\sigma}_{23}, \qquad (21)$$

where the nonlinearity functions which appear in the field variables describe the fact that, the constant couplings are transformed to intensity-dependent couplings as below,

$$g_1 \to g_1 f_1(\hat{n}_1.\hat{n}_2) = g_1 \left[1 - \varepsilon_1^2 \hat{n}_1 - \tfrac{\varepsilon_2^2}{2}(1+\hat{n}_2)\right],$$

$$g_2 \to g_2 f_2(\hat{n}_1.\hat{n}_2) = g_2 \left[1 - \varepsilon_2^2 \hat{n}_2 - \tfrac{\varepsilon_1^2}{2}(1+\hat{n}_1)\right], \qquad (22)$$

Also note that the parameters $\varepsilon_1 = \frac{g_1}{2\widetilde{\omega}_{12}}, \left(\widetilde{\omega}_{12} = \frac{E_1-E_2+\omega_f}{2}\right)$ and $\varepsilon_2 = \frac{g_2}{2\widetilde{\omega}_{23}}, \left(\widetilde{\omega}_{13} = \frac{E_2-E_3+\omega_{f'}}{2}\right), \gamma_1 = \frac{2\varepsilon_1^2 \widetilde{\omega}_{12}+\varepsilon_2^2 \widetilde{\omega}_{23}}{\omega_f(E_1-E_2-3\omega_f)} g_1, \gamma_2 = \frac{2\varepsilon_2^2 \widetilde{\omega}_{23}+\varepsilon_1^2 \widetilde{\omega}_{12}}{\omega_{f'}(E_2-E_3-3\omega_{f'})} g_2,$ and the parameters $\beta_1, \beta_2$ have been defined respectively in (8).

Now we want to repeat the procedure for a typical three-level atom in ladder (cascade) configuration as shown in figure 1(c). The permissable transitions are from the upper level $|1\rangle$ to the intermediate level $|2\rangle$ as well as from level $|2\rangle$ to the ground level $|3\rangle$. The Hamiltonian describing the interaction of ladder type atom with a two-mode field can be written as below, wherein we have kept the CRTs, too,

$$\hat{H} = \hat{H}_0 + \hat{H}_1 + \hat{H}_2, \quad (23)$$

where $\hat{H}_0$ has been introduced in Eq. (2) and

$$\hat{H}_1 = g_1(\hat{a}^\dagger \hat{\sigma}_{12} + \hat{a}\hat{\sigma}_{21}) + g_2(\hat{b}^\dagger \hat{\sigma}_{32} + \hat{\sigma}_{23}), \quad (24)$$

$$\hat{H}_2 = g_1(\hat{a}^\dagger \hat{\sigma}_{21} + \hat{a}\hat{\sigma}_{12}) + g_2(\hat{b}^\dagger \hat{\sigma}_{23} + \hat{b}\hat{\sigma}_{32}), \quad (25)$$

In this case also we apply three successive unitary transformations defined as below,

$$\hat{U}_1 = e^{\varepsilon_1(\hat{a}^\dagger \hat{\sigma}_{12} - \hat{a}\hat{\sigma}_{21}) + \varepsilon_2(\hat{b}^\dagger \hat{\sigma}_{23} - \hat{b}\hat{\sigma}_{32})}, \quad (26)$$

$$\hat{U}_2 = e^{\beta_1(\hat{a}^{\dagger 2} - \hat{a}^2)\hat{\sigma}_z^{12} + \beta_2(\hat{b}^{\dagger 2} - \hat{b}^2)\hat{\sigma}_z^{13}}, \quad (27)$$

$$\hat{U}_3 = e^{\gamma_1(\hat{a}^{\dagger 3}\hat{\sigma}_{12} - \hat{a}^3\hat{\sigma}_{21}) + \gamma_2(\hat{b}^{\dagger 3}\hat{\sigma}_{23} - \hat{b}^3\hat{\sigma}_{32})}, \quad (28)$$

which arrives one at the nonlinear (intensity-dependent) atom-field couplings in the dynamical Hamiltonian as below:

$$\hat{H}_{eff} \approx \hat{H}_0 + \hat{H}_1^D$$

$$+\varepsilon_1 g_1[(\hat{n}_1 + \tfrac{1}{2})\hat{\sigma}_z^{12} + \tfrac{\hat{\sigma}_{33}}{2} - \tfrac{1}{2}]$$

$$+\varepsilon_2 g_2 \left[\left(\hat{n}_2 + \tfrac{1}{2}\right)\hat{\sigma}_z^{23} + \tfrac{\hat{\sigma}_{11}}{2} - \tfrac{1}{2}\right], \quad (29)$$

with the definition,

$$\hat{H}_1^D = \hat{a} g_1 f_1(\hat{n}_1, \hat{n}_2)\hat{\sigma}_{12} + g_1 f_1(\hat{n}_1, \hat{n}_2)\hat{a}^\dagger \hat{\sigma}_{21}$$

$$+\hat{b} g_2 f_2(\hat{n}_1, \hat{n}_2)\hat{\sigma}_{23} + g_2 f_2(\hat{n}_1, \hat{n}_2)\hat{b}^\dagger \hat{\sigma}_{32}, \quad (30)$$

where the appeared nonlinearity functions are such that, in this Eq. the constant couplings in the initial Hamiltonian (3), (4) have been replaced by some intensity-dependent coupling functions as is explained in below,

$$g_1 \to g_1 f_1(\hat{n}_1, \hat{n}_2) = g_1 \left[1 - \varepsilon_1^2 \hat{n}_1 - \tfrac{\varepsilon_2^2}{2}\hat{n}_2\right],$$

$$g_2 \to g_2 f_2(\hat{n}_1, \hat{n}_2) = g_2 \left[1 - \tfrac{\varepsilon_1^2}{2}(1 + \hat{n}_1) - \varepsilon_2^2 \hat{n}_2\right], \quad (31)$$

Note that the parameters $\varepsilon_1(\varepsilon_2), \beta_1(\beta_2)$ and $\gamma_1(\gamma_2)$ have been defined in description of V atom.

**2.3 The eigenvalues of the Hamiltonians based on perturbation theory**

In this section, via working with the final form of the obtained transformed Hamiltonians in the previous section, the eigenvalues and eigenstates of the atom-field system will be obtained. We begin with the $\Lambda$-type atom, meanwhile, instead of addressing the Hamiltonian (11), it is more favorable to rewrite it as follows:

$$\hat{H}_{eff} = \omega_f \hat{n}_1 + \omega_{f'} \hat{n}_2 + \tfrac{1}{3}\omega_f \hat{\sigma}_z^{12} + \tfrac{1}{3}\omega_{f'} \hat{\sigma}_z^{13}$$

$$+ \tfrac{(2\Delta_{12} - \Delta_{13})}{3}\hat{\sigma}_z^{12} + \tfrac{(2\Delta_{13} - \Delta_{12})}{3}\hat{\sigma}_z^{13}$$

$$+\tfrac{1}{3}(E_1 + E_2 + E_3)(\hat{\sigma}_{11} + \hat{\sigma}_{22} + \hat{\sigma}_{33})$$

$$+\varepsilon_1 g_1(\hat{n}_1 \hat{\sigma}_z^{12} - \hat{\sigma}_{22}) + \varepsilon_2 g_2(\hat{n}_2 \hat{\sigma}_z^{13} - \hat{\sigma}_{33})$$

$$+\hat{a} g_1 f_1(\hat{n}_1, \hat{n}_2)\hat{\sigma}_{12} + g_1 f_1(\hat{n}_1, \hat{n}_2)\hat{a}^\dagger \hat{\sigma}_{12}$$

$$+ \hat{b} g_2 f_2(\hat{n}_1, \hat{n}_2)\hat{\sigma}_{13} + g_2 f_2(\hat{n}_1, \hat{n}_2)\hat{b}^\dagger \hat{\sigma}_{31}, \quad (32)$$

where the detuning parameters are defined as,

$$\Delta_{12} = E_1 - E_2 - \omega_f, \quad \Delta_{13} = E_1 - E_3 - \omega_{f'} \quad (33)$$

Based on the fact that $[\hat{N},\hat{H}_{eff}] = 0$, which leads one to the realization that $\hat{N}$ is a constant of motion, we can straightforwardly solve the eigenvalue equation associated with the Hamiltonian (29) for a constant quanta $N$. We consider the eigenvalue equation of effective Hamiltonian as follows:

$$\hat{H}_{eff}|\phi_L\rangle_{eff} = \mu_L|\phi_L\rangle_{eff}, L = 1,2,3, \quad (34)$$

Hence, for a fixed number of quanta $N = n + 1$, and in the unpaired atom-field states, $|1,n_1,n_2\rangle \equiv |1\rangle \otimes |n_1,n_2\rangle, |2,n_1+1,n_2\rangle \equiv |2\rangle \otimes |n_1+1,n_2\rangle, |3,n_1,n_2+1\rangle \equiv |3\rangle \otimes |n_1,n_2+1\rangle$, the eigenvalues can be estimated from the following matrix relation:

$$\begin{vmatrix} h_1 + h_2 + D_1 + D_2 - \mu_L & \sqrt{n_1+1}f_1(n_1+1,n_2) & \sqrt{n_2+1}f_2(n_1,n_2+1) \\ \sqrt{n_1+1}f_1(n_1+1,n_2) & h_2 - D_1 - \mu_L & 0 \\ \sqrt{n_2+1}f_2(n_1,n_2+1) & 0 & h_3 - D_2 - \mu_L \end{vmatrix} = 0, \quad (35)$$

where we have defined:

$$h_1 = \omega_f\left(n_1 + \frac{1}{3}\right) + \omega_{f'}\left(n_2 + \frac{1}{3}\right)g_1\hbar - \varepsilon_2 g_2,$$

$$h_2 = \omega_f\left(n_1 + \frac{1}{3}\right) + \omega_{f'}\left(n_2 + \frac{1}{3}\right) - \varepsilon_1 g_1 \hbar + \frac{(\Delta_{13}-\Delta_{12})}{3},$$

$$h_3 = \omega_f\left(n_1 + \frac{1}{3}\right) + \omega_{f'}\left(n_2 + \frac{1}{3}\right) - \varepsilon_2 g_2 \hbar + \frac{(\Delta_{12}-\Delta_{13})}{3},$$

$$D_1 = \frac{\Delta_{12}}{3} + \varepsilon_1 g_1(n_1 + 1),$$

$$D_2 = \frac{\Delta_{13}}{3} + \varepsilon_2 g_1(n_2 + 1). \quad (36)$$

Using some algebraic calculations, the above equation can be reduced to the following cubic equation,

$$\mu_L^3 + x_1 \mu_L^2 + x_2 \mu_L + x_3 = 0, \quad L = 1,2,3, \quad (37)$$

where $\mu_L$ is the eigenvalue that corresponds to $L$th eigenvector of $\hat{H}_{eff}$ and

$$x_1 = -(h_1 + h_2 + h_3),$$

$$x_2 = -(n_1 + 1)f_1^2(n_1+1,n_2) \\ -(n_2 + 1)f_2^2(n_1,n_2+1) + h_1 h_2 \\ + h_1 h_3 + h_2 h_3 + (h_2 - h_1)D_1 \\ + (h_3 - h_1)D_2 - D_1 D_2 \\ - D_1^2 - D_2^2,$$

$$x_3 = (n_1+1)f_1^2(n_1+1,n_2)h_3 \\ +(n_2+1)f_2^2(n_1,n_2+1)h_2 \\ -h_1 h_2 h_3 + h_1 h_3 D_1 - h_2 h_3 D_1 \\ +h_1 h_2 D_2 - h_2 h_3 D_2 + h_3 D_1^2 \\ +h_2 D_2^2 - D_2 D_1^2 - D_1 D_2^2 + \\ (h_2 + h_3 - h_1)D_1 D_2 \\ -(n_1+1)f_1^2(n_1+1,n_2)D_2 \\ -(n_2+1)f_2^2(n_1,n_2+1). \quad (38)$$

To solve the cubic algebraic equation, we used the Cardano formula [36], according to which the eigenvalues of $\hat{H}_{eff}$ are given by:

$$\mu_L(n) = -\frac{1}{3}x_1 + \frac{2}{3}\sqrt{x_1^2 - 3x_2}\cos\left[\theta(n) + \frac{2}{3}(L-1)\pi\right], \quad (39)$$

$$\theta(n) = \frac{1}{3}\cos^{-1}\left(\frac{9x_1 x_2 - 2x_1^3 - 27x_3}{2(x_1^2 - 3x_2)^{\frac{3}{2}}}\right). \quad (40)$$

Also for the corresponding eigenvectors one finds:

$$|\phi_l\rangle_{eff} = a_l(n,m)|1,n_1,n_2\rangle + b_l(n,m)|2,n_1+1,n_2\rangle + c_l(n,m)|3,n_1,n_2+1\rangle, \quad (41)$$

$$\begin{pmatrix} h_1 + h_2 + D_1 + D_2 - \mu_L & \sqrt{n_1+1}f_1(n_1+1,n_2) & \sqrt{n_2+1}f_2(n_1,n_2+1) \\ \sqrt{n_1+1}f_1(n_1+1,n_2) & h_2 - D_1 - \mu_L & 0 \\ \sqrt{n_2+1}f_2(n_1,n_2+1) & 0 & h_3 - D_2 - \mu_L \end{pmatrix}$$

$$\begin{pmatrix} a_l \\ b_l \\ c_l \end{pmatrix} = \mu_L \begin{pmatrix} a_l \\ b_l \\ c_l \end{pmatrix}, \quad (42)$$

By combining the above equations and the probability conservation relation:

$$|a_l(n_1,n_2)|^2 + |b_l(n_1,n_2)|^2 + |c_l(n_1,n_2)|^2 = 1, \quad (43)$$

expansion coefficients are obtained as follows:

$$|a_l(n_1,n_2)| = \left(1 + \frac{(n_1+1)f_1^2(n_1+1,n_2)}{(\mu_L(n_1,n_2)-h_2+D_1)^2} + \frac{(n_2+1)f_2^2(n_1,n_2+1)}{(\mu_L(n_1,n_2)-h_3+D_2)^2}\right)^{-\frac{1}{2}}, \quad |b_l(n_1,n_2)| = \frac{\sqrt{n_1+1}f_1(n_1+1,n_2)}{\mu_L(n_1,n_2)-h_2+D_1}|a_l(n_1,n_2)|,$$

$$|c_l(n_1.n_2)| = \frac{\sqrt{n_1+1}f_2(n_1,n_2+1)}{\mu_L(n_1.n_2)-h_3+D_2}|a_l(n_1,n_2)|, \quad (44)$$

Now we consider the eigenvalue equation for $\widehat{H}$, which is defined by Eq. (1) in the form:

$$\widehat{H}|\phi_L\rangle = \mu_L|\phi_L\rangle, \qquad L = 1,2,3. \quad (45)$$

Keeping in mind the fundamentals of perturbation theory, it seems reasonable that the eigenvalues $\mu_l$ are approximated to the eigenvalues $\mu_L$. Also, according to the transformations which are applied in the previous section, the following relationship is established between the eigenvectors of $\widehat{H}$ and $\widehat{H}_{eff}$:

$$|\phi_L\rangle = U_1^\dagger U_2^\dagger U_3^\dagger |\phi_L\rangle_{eff}, \quad (46)$$

$$|\phi_L\rangle \approx 1 - \varepsilon_1(\hat{a}^\dagger\hat{\sigma}_{12} - \hat{a}\hat{\sigma}_{21}) - \varepsilon_2(\hat{b}^\dagger\hat{\sigma}_{13} - \hat{b}\hat{\sigma}_{31})|\phi_l\rangle_{eff} = a_l(n_1,n_2)|1,n_1,n_2\rangle + b_l(n_1,n_2)|2,n_1+1,n_2\rangle + c_l(n_1,n_2)|3,n_1,n_2+1\rangle - \varepsilon_1\sqrt{n_1}a_l(n_1,n_2)|2,n_1-1,n_2\rangle - \sqrt{n_1+2}b_l(n_1,n_2)|1,n_1+2,n_2+1\rangle - \varepsilon_2\sqrt{n_2}a_l(n_1,n_2)|3,n_1-1,n_2\rangle - \sqrt{n_2+2}c_l(n_1,n_2)|1,n_1+1,n_2+2\rangle, \quad (47)$$

**2.3 The eigenvalues of the Hamiltonians based on perturbation theory**

In this subsection, we first assume that the atom is prepared in the state basis 1 at the onset of interaction and the field is a bimodal coherent state. So, the initial state of the atom-field system reads as,

$$|\psi(0)\rangle = \sum_{n_1=0}^{\infty}\sum_{n_2=0}^{\infty} P_{n_1}P_{n_2}|1,n_1,n_2\rangle,$$

$$P_{n_i} = \frac{e^{\frac{-|\alpha|^2}{2}}\alpha^{n_i}}{\sqrt{n_i!}}, \quad i = n_1,n_2, \quad (48)$$

where $|\alpha|^2$ is the average number of photons in the field at $t = 0$. One can easily obtain the atom-field state vector of the system at time $t$ by applying the time evolution operator onto the assumed initial state:

$$|\psi(t)\rangle = e^{-\frac{i}{\hbar}\widehat{H}t}|\psi(0)\rangle = \sum_{n_1=0}^{\infty}\sum_{n_2=0}^{\infty}\sum_{L=1}^{3} e^{-\frac{i\mu_L(n_1,n_2)t}{\hbar}}\lambda_l(n_1,n_2)|\phi_l\rangle, \quad (49)$$

$$\lambda_l(n_1,n_2) = \langle\phi_L|\psi(0)\rangle = a_l(n_1,n_2)P_{n_1}P_{n_2} - \varepsilon_1 b_l(n_1,n_2)\sqrt{n_1+2}P_{n_1+2}P_{n_2+1} - \varepsilon_2 c_l(n_1,n_2)\sqrt{n_2+2}P_{n_1+1}P_{n_2+2}, \quad (50)$$

the state vector at any time up to the first-order approximation may be obtained in the following form:

$$|\psi(t)\rangle = \sum_{n_1=0}^{\infty}\sum_{n_2=0}^{\infty} (A_{n_1,n_2,t}|1,n_1,n_2\rangle + B_{n_1+1,n_2,t}|2,n_1+1,n_2\rangle + C_{n_1,n_2+1,t}|3,n_1,n_2+1\rangle), \quad (51)$$

where the probability amplitudes $A_{n_1,n_2,t}$, $B_{n_1+1,n_2,t}$ and $C_{n_1,n_2+1,t}$ in Eq. (44) have been determined by the following relations:

$$A_{n_1,n_2,t} = \sum_{l=1}^{3} e^{-i\mu_L(n_1,n_2)t}(a_l(n_1,n_2)^2 P_{n_1}P_{n_2} - \varepsilon_1 b_l(n_1,n_2)a_l(n_1,n_2)\sqrt{n_1+2}P_{n_1+2}P_{n_2} - \varepsilon_2 a_l(n_1,n_2)c_l(n_1,n_2)\sqrt{n_2+2}P_{n_1}P_{n_2+2}) - e^{-i\mu_L(n_1-2,n_2)t}\varepsilon_1 b_l(n_1-2,n_2)a_l(n_1-2,n_2)\sqrt{n_1}P_{n_1-2}P_{n_2} + e^{-i\mu_L(n_1,n_2-2)t}\varepsilon_2\sqrt{n_2}a_l(n_1,n_2-2)c_l(n_1,n_2-2)P_{n_1}P_{n_2-2}, \quad (52)$$

$$B_{n_1+1,n_2,t} = \sum_{l=1}^{3} e^{-i\mu_L(n_1,n_2)t}(a_l(n_1,n_2)b_l(n_1,n_2)P_{n_1}P_{n_2} - \varepsilon_1 b_l(n_1,n_2)^2\sqrt{n_1+2}P_{n_1+2}P_{n_2} - \varepsilon_2 b_l(n_1,n_2)c_l(n_1,n_2)\sqrt{n_2+2}P_{n_1}P_{n_2+2}) + e^{-i\mu_L(n_1+2,n_2)t}\varepsilon_1 a_l(n_1-2,n_2)^2\sqrt{n_1+2}P_{n_1+2}P_{n_2} -$$

$e^{-i\mu_L(n_1,n_2+2)t}\varepsilon_2\sqrt{n_2+2}a_l(n_1,n_2-2)^2 P_{n_1}P_{n_2+2},$    (53)

$C_{n_1,n_2+1,t} =$
$\sum_{l=1}^{3} e^{-i\mu_L(n_1,n_2)t}(a_l(n_1,n_2)c_l(n_1,n_2)P_{n_1}P_{n_2}$

$-\varepsilon_1 b_l(n_1,n_2)c_l(n_1,n_2)\sqrt{n_1+2}P_{n_1+2}P_{n_2}$

$-\varepsilon_2 c_l(n_1,n_2)^2\sqrt{n_2+2}P_{n_1}P_{n_2+2})$

$+e^{-i\mu_L(n_1+2,n_2)t}\varepsilon_1 a_l(n_1-2,n_2)^2\sqrt{n_1+2}P_{n_1+2}P_{n_2}$

$-e^{-i\mu_L(n_1,n_2+2)t}\varepsilon_2\sqrt{n_2+2}a_l(n_1,n_2-2)^2 P_{n_1}P_{n_2+2},$    (54)

It should be noted that due the approximations used in the processes of solution, normalization of the state vector should be corrected. In order to be certain about this fact, we rewrite the system state as:

$|\psi(t)\rangle_{norm} = N(t)|\psi(t)\rangle,$    (55)

where

$N(t) = \langle\psi(t)|\psi(t)\rangle^{-\frac{1}{2}} =$
$\left(\sum_{n_1=0}^{\infty}\sum_{n_2=0}^{\infty}(|A_{n_1,n_2,t}|^2 + |B_{n_1+1,n_2,t}|^2 + |C_{n_1,n_2+1,t}|^2)\right)^{-\frac{1}{2}}.$    (56)

## III. NONCLASSICAL PROPERTIES OF THE SYSTEM AND THE EFFECTS OF CRTs

Searching for nonclassical properties of the atom-field interactions is of enough interest [37-39]. In this section we use a few nonclassical features including the population inversion of atomic levels and photon statistics of the quantized field, with and without RWA. For this purpose, we plot the latter quantities in terms of time for constant amounts of $\frac{\Delta_{12}}{\omega_f} = 0.2$, $\frac{\Delta_{13}}{\omega_{f'}} = 0.28$ and $|\alpha|^2 = 25$. Then, we compare the results from the plotted figures to investigate the effect of the CRTs on the time evolution of the dynamical properties of the state of the system.

### 3.1 Population inversion of atoms

Population inversion concerning the considered atomic systems indicates difference between higher level population(s) with the total population of the lower level(s), i.e,

$W(t) = N(t)^2(\sum_{n_1=0}^{\infty}\sum_{n_2=0}^{\infty}(|A_{n_1,n_2,t}|^2 - |B_{n_1+1,n_2,t}|^2 - |C_{n_1,n_2+1,t}|^2)).$    (57)

According to figure 2, for non-resonance conditions, the evolution of the atomic population inversion possesses collapses and revivals during the interaction in all cases typically. In more detail, while in the presence of RWA, very clear collapse-revivals may be observed, absence of RWA the patterns.

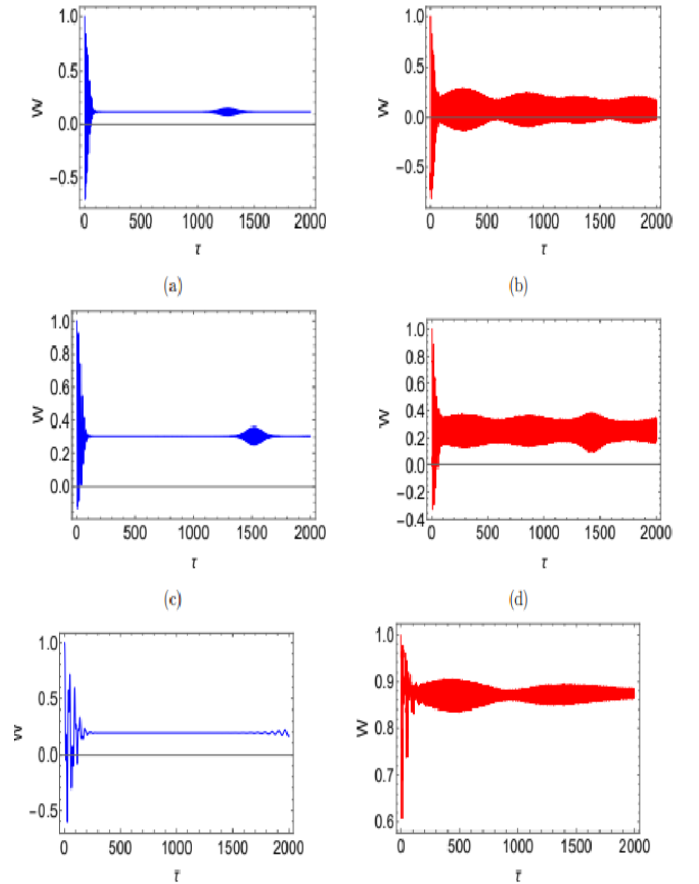

Fig. 2 The time evolution of the atomic population inversion versus the scaled time $\tau = \omega_f t$. Here, we considered $\frac{\Delta_{12}}{\omega_f} = 0.2, \frac{\Delta_{13}}{\omega_{f'}} = 0.28$, $|\alpha|^2 = 25$ and $\frac{g_1}{\omega_f} == 0.01$, $\frac{g_2}{\omega_{f'}} = 0.04$. The left curves correspond to the presence of RWA ($\varepsilon_1 = \varepsilon_2 = 0$) and the right curves

are obtained in the absence of RWA ($\varepsilon_1 = \frac{1}{55}, \varepsilon_2 = \frac{1}{57}$). Top plots are related to the $\Lambda$, middled plots for the $V$ and low plots are for the ladder atom.

changes and the sharp collapse-revivals can no longer be observed. All we may state is that typical collapse-revival may be observed with the CRTs. It can be stated that in the latter case, sharp and low-amplitude fluctuations are affected by virtual transitions caused by CRTs.

### 3.2 Photon statistics of the fields: Mandel parameter

In order to examine the field photon statistics during the interaction with each of the three-level atoms, the Mandel parameter is used as follows:

$$Q = \frac{\langle \hat{n}^2 \rangle - \langle \hat{n} \rangle^2}{\langle \hat{n} \rangle} - 1, \quad (58)$$

If for a field state, $-1 \leq Q < 0$, the photon statistics are sub-Poissonian (non-classical light), if $Q = 0$, the field possesses Poissonian distribution (standard coherent light) and if $Q > 0$ the field has super-Poissonian distribution (classical light). The temporal evolution of the Mandel parameter in terms of the scaled time is shown in figure 3, where we consider the atomic configurations and required parameters for plotting similar to figure 2. Our numerical calculations show that Mandel parameter also involves fluctuations in the form of collapse-revival phenomenon. But in the presence of RWA (right figures) one observe very clear collapse-revival patterns, while in the presence of RWA (left figures), the patterns will be changed and we have only, say typical collapse-revivals. These effects may also be attributed to the virtual transitions due to the presence of CRTs in the system Hamiltonian.

### IV. SUMMARY AND CONCLUSIONS

In this paper, based on the perturbation theory approach, we considered the quantum interaction between a three-level atom with a two-mode quantized field without RWA. We have solved analytically the Hamiltonian interaction corresponding to all types of three-level atoms ($\Lambda, V, \Xi$). At first, the contribution of CRTs in the first-order approximation appears as intensity-dependent displacement in the energy of the atom-field levels and in the second-order approximation two intensity-dependent functions appear instead of coupling constants. By solving the effective Hamiltonian dynamics, we have obtained the associated eigenvalues and the eigenstates, and through them, we have arrived at the approximate eigenvalues and the eigenstates Hamiltonian of the interaction system. Since in solving process, the second-order terms of the parameters of the perturbation ($\varepsilon_1, \varepsilon_2$) have been omitted, our approach is valid only for small values of these parameters. In the continuation, we have investigated the temporal evolution of the system in the presence and absence and presence of RWA by using the state vector obtained through our calculations in a non-resonant condition. As we show through the numerical results, the population inversion and Mandel parameter both have collapse and revival patterns. However, while in the absence of RWA we observe a typical collapse-revival, in the presence of RWA we have a clarified pattern of collapse-revival phenomenon. In fact the presence of CRTs causes that the apparent patterns of collapse-revivals being destroyed. In general, the observed effects can be attributed to virtual (photons) transitions created by CRTs in the atom-field system Hamiltonian. Altogether, what we can say about the nonclassicality signs is that in both cases, either with or without CRTs and in all configurations of atoms the considered systems possess such nonclassicalities clearly.

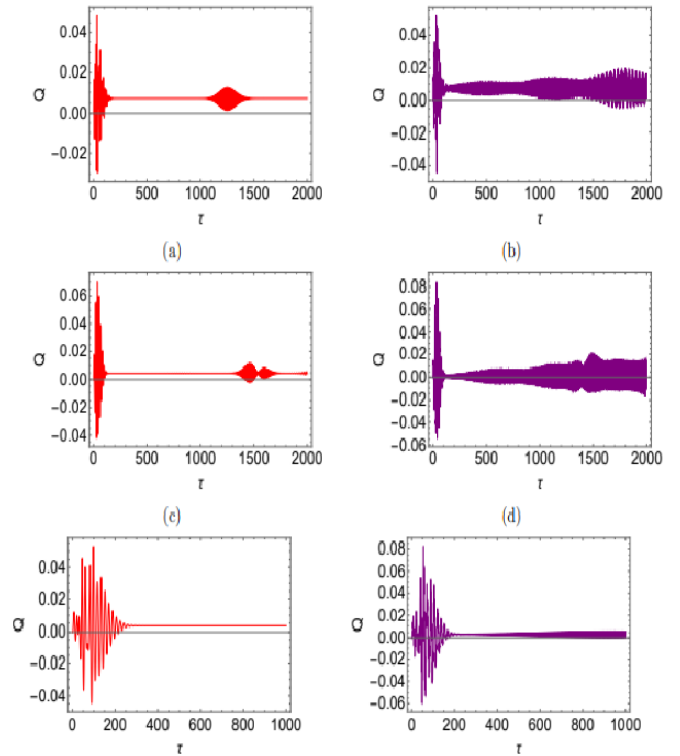

Fig. 3 The time evolution of Mandel parameter for chosen parameters similar to figure 2.

**APPENDIX**

For ladder atom

$$\hat{H}_{eff} = \omega_f \hat{n}_1 + \omega_{f'} \hat{n}_2 + \omega_f \hat{\sigma}_z^{12} + \omega_{f'} \hat{\sigma}_z^{23} + \frac{(2\Delta_{12}+\Delta_{23})}{3}\hat{\sigma}_z^{12} + \frac{(2\Delta_{23}+\Delta_{12})}{3}\hat{\sigma}_z^{23} + \frac{1}{3}(E_1+E_2+E_3)(\hat{\sigma}_{11}+\hat{\sigma}_{22}+\hat{\sigma}_{33}) + \varepsilon_1 g_1(\hat{n}_1\hat{\sigma}_z^{12}-\hat{\sigma}_{22}) + \varepsilon_2 g_2(\hat{n}_2\hat{\sigma}_z^{23}-\hat{\sigma}_{33}) + \hat{a}g_1 f_1(\hat{n}_1,\hat{n}_2)\hat{\sigma}_{12} + g_1 f_1(\hat{n}_1,\hat{n}_2)\hat{a}^\dagger\hat{\sigma}_{21} + \hat{b}g_2 f_2(\hat{n}_1,\hat{n}_2)\hat{\sigma}_{23} + g_2 f_2(\hat{n}_1,\hat{n}_2)\hat{b}^\dagger\hat{\sigma}_{32}, \quad (59)$$

$$\Delta_{12} = E_1 - E_2 - \omega_f, \Delta_{23} = E_2 - E_3 - \omega_{f'}, \quad (60)$$

Hence, for a fixed quantum number $N = n + 1$, and in unpaired atom-field states, $|1,n_1,n_2\rangle \equiv |1\rangle \otimes, |2,n_1+1,n_2\rangle \equiv |2\rangle \otimes |n_1+1,n_2\rangle, |3,n_1,n_2+1\rangle \equiv |3\rangle \otimes |n_1+1,n_2+1\rangle$, the eigenvalues can be estimated from the following matrix relation:

$$\begin{vmatrix} h_1-\mu_L & \sqrt{n_1+1}f_1(n_1+1,n_2) & 0 \\ \sqrt{n_1+1}f_1(n_1+1,n_2) & h_2-\mu_L & \sqrt{n_2+1}f_2(n_1,n_2+1) \\ 0 & \sqrt{n_2+1}f_2(n_1,n_2+1) & h_3-\mu_L \end{vmatrix} = 0, \quad (61)$$

where we have defined:

$$h_1 = \omega_f\left(n_1+\frac{1}{2}\right) + \omega_{f'}\left(n_2+\frac{1}{2}\right) + \varepsilon_1 g_1 n_1 + \frac{(2\Delta_{12}+\Delta_{23})}{3},$$

$$h_2 = \omega_f\left(n_1+\frac{1}{2}\right) + \omega_{f'}\left(n_2+\frac{1}{2}\right) - \varepsilon_1 g_1 - \varepsilon_1 g_1 n_1 + \varepsilon_2 g_2 n_2 + \frac{(\Delta_{13}-\Delta_{12})}{3},$$

$$h_3 = \omega_f\left(n_1+\frac{1}{2}\right) + \omega_{f'}\left(n_2+\frac{1}{2}\right) - \varepsilon_2 g_2 - \varepsilon_2 g_2 n_2 - \frac{(\Delta_{12}+2\Delta_{23})}{3}. \quad (62)$$

Using some algebraic calculations, the above equation is reduced to the following algebraic cubic equation:

$$\mu_L^3 + x_1\mu_L^2 + x_2\mu_L + x_3, \quad L = 1,2,3 \quad (63)$$

where $\mu_L$ is the eigenvalue corresponding to the $L$th eigenvector of $\hat{H}_{eff}$ and

$$x_1 = -(h_1 + h_2 + h_3), \quad (64)$$

$$x_2 = -(n_1+1)f_1^2(n_1+1,n_2) - (n_2+1)f_2^2(n_1,n_2+1) + h_1 h_2 + h_1 h_3 + h_2 h_3,$$

$$x_3 = (n_1+1)f_1^2(n_1+1,n_2)h_3 + (n_2+1)f_2^2(n_1.n_2+1)h_1 - h_1 h_2 h_3,$$

$$|\phi_l\rangle_{eff} = a_l(n_1,n_2)|1,n_1,n_2\rangle + b_l(n_1,n_2)|2,n_1+1,n_2\rangle + c_l(n_1,n_2)|3,n_1+1,n_2+1\rangle, \quad (65)$$

$$\begin{pmatrix} h_1-\mu_L & \sqrt{n_1+1}f_1(n_1+1,n_2) & 0 \\ \sqrt{n_1+1}f_1(n_1+1,n_2) & h_2-\mu_L & \sqrt{n_2+1}f_2(n_1,n_2+1) \\ 0 & \sqrt{n_2+1}f_2(n_1,n_2+1) & h_3-\mu_L \end{pmatrix}$$

$$\begin{pmatrix} a_l \\ b_l \\ c_l \end{pmatrix} = \mu_L \begin{pmatrix} a_l \\ b_l \\ c_l \end{pmatrix}, \quad (66)$$

with the following expansion coefficients:

$$|a_l(n_1,n_2)| = \frac{\sqrt{n_1+1}f_1(n_1+1,n_2)}{\mu_L(n_1,n_2)-h_1}|b_l(n_1,n_2)|,$$

$$|b_l(n_1,n_2)| = \left(1 + \frac{(n_1+1)f_1^2(n_1+1,n_2)}{(\mu_L(n_1,n_2)-h_1)^2} + \frac{(n_2+1)f_2^2(n_1,n_2+1)}{(\mu_L(n_1,n_2)-h_3)^2}\right)^{-\frac{1}{2}},$$

$$|c_l(n_1,n_2)| = \frac{\sqrt{n_2+1}f_2(n_1,n_2+1)}{\mu_L(n_1,n_2)-h_3}|b_l(n_1,n_2)|, \qquad (67)$$

Now, we consider the eigenvalue equation for $\hat{H}$, which is defined by Eq. (1) in the form:

$$\hat{H}|\phi_L\rangle = \mu_L|\phi_L\rangle, \qquad L = 1,2,3, \qquad (68)$$

$$|\phi_L\rangle = U_1^\dagger U_2^\dagger U_3^\dagger |\phi_L\rangle_{eff}, \qquad (69)$$

$$|\phi_L\rangle \approx 1 - \varepsilon_1(\hat{a}^\dagger\hat{\sigma}_{12} - \hat{a}\hat{\sigma}_{21}) - \varepsilon_2(\hat{b}^\dagger\hat{\sigma}_{23} - \hat{b}\hat{\sigma}_{32})|\phi_l\rangle_{eff},$$

$$= a_l(n_1,n_2)|1,n_1,n_2\rangle$$
$$+ b_l(n_1,n_2)|2,n_1+1,n_2\rangle$$
$$+ c_l(n_1,n_2)|3,n_1+1,n_2+1\rangle$$
$$- \varepsilon_1\sqrt{n_1+2}b_l(n_1,n_2)|1,n_1+2,n_2\rangle$$
$$+ \varepsilon_1\sqrt{n_1}a_l(n_1,n_2)|2,n_1-1,n_2\rangle$$
$$- \varepsilon_2\sqrt{n_2+2}c_l(n_1,n_2)|2,n_1+1,n_2+2\rangle$$
$$+ \varepsilon_2\sqrt{n_2}bl(n_1,n_2)|3,n_1+1,n_2+1\rangle, \qquad (70)$$

$$\lambda_l(n_1,n_2) = \langle\phi_L|\psi(0)\rangle = a_l(n_1,n_2)P_{n_1}P_{n_2} - \varepsilon_1 b_l(n_1,n_2)\sqrt{n_1+2}P_{n_1+2}P_{n_2}, \qquad (71)$$

The state vector of the system, at any time, up to the first-order approximation may be obtained in the following form:

$$|\psi(t)\rangle = \sum_{n_1=0}^{\infty}\sum_{n_2=0}^{\infty} (A_{n_1.n_2.t}|1,n_1,n_2\rangle$$
$$+ B_{n_1+1.n_2.t}|2,n_1+1\rangle$$
$$+ C_{n_1.n_2+1.t}|3,n_1+1,n_2+1\rangle, \qquad (72)$$

$$A_{n_1.n_2.t} = \sum_{l=1}^{3} e^{-i\mu_L(n_1,n_2)t}(a_l(n_1,n_2)^2 P_{n_1}P_{n_2}$$
$$- \varepsilon_1 b_l(n_1,n_2)a_l(n_1,n_2)\sqrt{n_1+2}P_{n_1+2}P_{n_2}$$
$$- e^{-i\mu_L(n_1-2,n_2)t}\varepsilon_1 b_l(n_1-2,n_2)a_l(n_1-2,n_2)\sqrt{n_1}P_{n_1-2}P_{n_2}, \qquad (73)$$

$$B_{n_1+1,n_2,t} = \sum_{l=1}^{3} e^{-i\mu_L(n_1,n_2)t}(a_l(n_1,n_2)b_l(n_1,n_2)P_{n_1}P_{n_2}$$
$$- \varepsilon_1 b_l(n_1,n_2)^2\sqrt{n_1+2}P_{n_1+2}P_{n_2}$$
$$- e^{-i\mu_L(n_1+2,n_2)t}\varepsilon_2 a_l(n_1,n_2-2)c_l(n_1,n_2-2)\sqrt{n_2}P_{n_1}P_{n_2-2} +$$
$$e^{-i\mu_L(n_1+2,n_2)t}\varepsilon_2\sqrt{n_2+2}a_l(n_1+2,n_2)^2 P_{n_1}P_{n_2+2}, \qquad (74)$$

$$C_{n_1+1,n_2+1,t} = \sum_{l=1}^{3} e^{-i\mu_L(n_1,n_2)t}(a_l(n_1,n_2)c_l(n_1,n_2)P_{n_1}P_{n_2}$$
$$- \varepsilon_1 b_l(n_1,n_2)c_l(n_1,n_2)\sqrt{n_1+2}P_{n_1+2}P_{n_2})$$
$$+ e^{-i\mu_L(n_1,n_2+2)t}\varepsilon_2\sqrt{n_2+2}a_l(n_1,n_2+2)^2 P_{n_1}P_{n_2+2}, \qquad (75)$$

Similarly, for $V-$type atom we have,

$$\hat{H}_{eff} = \omega_f \hat{n}_1 + \omega_{f'}\hat{n}_2 + \frac{1}{3}\omega_f \hat{\sigma}_z^{13} + \frac{1}{3}\omega_{f'}\hat{\sigma}_z^{23}$$
$$+ \frac{(2\Delta_{13}-\Delta_{23})}{3}\hat{\sigma}_z^{13} + \frac{(2\Delta_{23}-\Delta_{13})}{3}\hat{\sigma}_z^{23}$$
$$+ \frac{1}{3}(E_1+E_2+E_3)(\hat{\sigma}_{11}+\hat{\sigma}_{22}+\hat{\sigma}_{33})$$
$$+ \varepsilon_1 g_1(\hat{n}_1\hat{\sigma}_z^{13} - \hat{\sigma}_{33}) + \varepsilon_2 g_2(\hat{n}_2\hat{\sigma}_z^{23} - \hat{\sigma}_{33})$$
$$+ \hat{a}g_1 f_1(\hat{n}_1,\hat{n}_2)\hat{\sigma}_{13} + g_1 f_1(\hat{n}_1,\hat{n}_2)\hat{a}^\dagger\hat{\sigma}_{31}$$
$$+ \hat{b}g_2 f_2(\hat{n}_1,\hat{n}_2)\hat{\sigma}_{23} + g_2 f_2(\hat{n}_1,\hat{n}_2)\hat{b}^\dagger\hat{\sigma}_{32},$$
$$\Delta_{12} = E_1 - E_2 - \omega_f, \quad \Delta_{23} = E_2 - E_3 - \omega_{f'}, \qquad (76)$$

Hence, for a fixed quantum number $N = n + 1$, and in unpaired atom-field states,

$|1,n_1,n_2\rangle \equiv |1\rangle \otimes |n_1,n_2\rangle$, $\frac{1}{\sqrt{2}}(|2,n_1+1,n_2\rangle + |2,n_1,n_2+1\rangle) \equiv \frac{1}{\sqrt{2}}|2\rangle \otimes (|n_1+1+n_2+1\rangle)$, $|3,n_1,n_2\rangle \equiv |3\rangle \otimes |n_1,n_2\rangle$, the eigenvalues can be estimated from the following matrix relation:

$$\begin{vmatrix} h_1 - \mu_L & 0 & \frac{1}{\sqrt{2}}\sqrt{n_2+1}f_2(n_1,n_2+1) \\ 0 & h_2 - \mu_L & \frac{1}{\sqrt{2}}\sqrt{n_1+1}f_1(n_1+1,n_2) \\ \frac{1}{\sqrt{2}}\sqrt{n_2+1}f_2(n_1,n_2+1) & \frac{1}{\sqrt{2}}\sqrt{n_1+1}f_1(n_1+1,n_2) & h_3 - \mu_L \end{vmatrix} = 0, \quad (77)$$

where we have defined:

$$h_1 = \omega_f n_1 + \omega_{f'}\left(n_2 + \frac{1}{3}\right) + \varepsilon_1 g_1 n_1 + \frac{(2\Delta_{13} + \Delta_{23})}{3},$$

$$h_2 = \omega_f n_1 + \omega_{f'}\left(n_2 + \frac{1}{3}\right) - \varepsilon_1 g_1 - \varepsilon_1 g_1 n_1 + \frac{(\Delta_{23} - \Delta_{13})}{3},$$

$$h_3 = \omega_f\left(n_1 + \frac{1}{2}\right) + \omega_{f'}\left(n_2 + \frac{1}{2}\right) - \varepsilon_2 g_2 - \varepsilon_2 g_2 n_2 - \frac{(\Delta_{12} + \Delta_{23})}{3}, \quad (78)$$

Ding algebraic calculations, the above equation is reduced to the following algebraic cubic equation:

$$\mu_L^3 + x_1\mu_L^2 + x_2\mu_L + x_3 = 0, \quad L = 1,2,3 \quad (79)$$

where $\mu_L$ is the eigenvalue that corresponds to $L$th eigenvector of $\hat{H}_{eff}$ and

$$x_1 = -(h_1 + h_2 + h_3),$$

$$x_2 = -(n_1+1)f_1^2(n_1+1,n_2) - (n_2+1)f_2^2(n_1,n_2+1) + h_1h_2 + h_1h_3 + h_2h_3,$$

$$x_3 = (n_1+1)f_1^2(n_1+1,n_2)h_3 + (n_2+1)f_2^2(n_1,n_2+1)h_1 - h_1h_2h_3, \quad (80)$$

$$|\phi_l\rangle_{eff} = a_l(n_1,n_2)|1,n_1,n_2\rangle + b_l(n_1,n_2)\frac{1}{\sqrt{2}}(|2,n_1+1,n_2\rangle + |2,n_1,n_2+1\rangle) + c_l(n_1,n_2)|3,n_1,n_2\rangle \quad (81)$$

$$\begin{pmatrix} h_1 - \mu_L & 0 & \frac{1}{\sqrt{2}}\sqrt{n_2+1}f_2(n_1,n_2+1) \\ 0 & h_2 - \mu_L & \frac{1}{\sqrt{2}}\sqrt{n_1+1}f_1(n_1+1,n_2) \\ \frac{1}{\sqrt{2}}\sqrt{n_2+1}f_2(n_1,n_2+1) & \frac{1}{\sqrt{2}}\sqrt{n_1+1}f_1(n_1+1,n_2) & h_3 - \mu_L \end{pmatrix}$$

$$\begin{pmatrix} a_l \\ b_l \\ c_l \end{pmatrix} = \mu_L \begin{pmatrix} a_l \\ b_l \\ c_l \end{pmatrix}, \quad (82)$$

with the following expansion coefficients,

$$|a_l(n_1,n_2)| = \frac{\sqrt{n_1+1}f_1(n_1+1,n_2)}{\sqrt{2}(\mu_L(n_1,n_2)-h_1)}|b_l(n_1,n_2)|,$$

$$|b_l(n_1,n_2)| = \left(1 + \frac{(n_1+1)_1^2(n_1+1,n_2)}{2(\mu_L(n_1,n_2)-h_1)^2} + \frac{(n_2+1)f_2^2(n_1,n_2+1)}{2(\mu_L(n_1,n_2)-h_3)^2}\right)^{-\frac{1}{2}},$$

$$|c_l(n_1,n_2)| = \frac{\sqrt{n_2+1}f_2(n_1,n_2+1)}{\sqrt{2}(\mu_L(n_1,n_2)-h_3)}|b_l(n_1,n_2)|. \quad (83)$$

Now, we consider the eigenvalue equation for $\hat{H}$, which is defined by equation (1) in the form:

$$\hat{H}|\phi_L\rangle = \mu_L|\phi_L\rangle, \quad L = 1,2,3, \quad (84)$$

$$|\phi_L\rangle = U_1^\dagger U_2^\dagger U_3^\dagger |\phi_L\rangle_{eff}, \quad (85)$$

$$\approx 1 - \varepsilon_1(\hat{a}^\dagger\hat{\sigma}_{12} - \hat{a}\hat{\sigma}_{21}) - \varepsilon_2(\hat{b}^\dagger\hat{\sigma}_{32} - \hat{b}\hat{\sigma}_{23})|\phi_l\rangle_{eff}$$

$$= a_l(n_1,n_2)|1,n_1,n_2\rangle + b_l(n_1,n_2)|2,n_1+1,n_2\rangle + c_l(n_1,n_2)|3,n_1+1,n_2+1\rangle$$

$$-\varepsilon_1\sqrt{n_1+2}b_l(n_1,n_2)|1,n_1+2,n_2\rangle$$

$$+\varepsilon_1\sqrt{n_1}a_l(n_1,n_2)|2,n_1-1,n_2\rangle$$

$$-\varepsilon_2\sqrt{n_2+2}c_l(n_1,n_2)|2,n_1+1,n_2+2\rangle$$

$$+\varepsilon_2\sqrt{n_2}bl(n_1,n_2)|3,n_1+1,n_2+1\rangle, \quad (86)$$

$$\lambda_l(n_1,n_2) = \langle\phi_L|\psi(0)\rangle = a_l(n_1,n_2)P_{n_1}P_{n_2}$$
$$-\varepsilon_1 b_l(n_1,n_2)\sqrt{n_1+2}P_{n_1+2}P_{n_2}, \quad (87)$$

The state vector of the system at any time $t$ up to the first-order approximation may be obtained in the following form:

$$|\psi(t)\rangle = \sum_{n_1=0}^{\infty}\sum_{n_2=0}^{\infty}(A_{n_1,n_2,t}|1,n_1,n_2\rangle$$

$$+B_{n_1+1,n_2,t}|2,n_1+1,n_2\rangle$$

$$+C_{n_1,n_2+1,t}|3,n_1+1,n_2+1\rangle, \quad (88)$$

$$A_{n_1,n_2,t} =$$
$$\sum_{l=1}^{3}e^{-i\mu_L(n_1,n_2)t}(a_l(n_1,n_2)^2 P_{n_1}P_{n_2}$$

$$-\varepsilon_1 b_l(n_1,n_2)a_l(n_1,n_2)\sqrt{n_1+2}P_{n_1+2}P_{n_2}$$

$$-e^{-i\mu_L(n_1-2,n_2)t}\varepsilon_1 b_l(n_1-2,n_2)a_l(n_1-2,n_2)\sqrt{n_1}P_{n_1-2}P_{n_2}, \quad (89)$$

$$B_{n_1+1,n_2+1,t} = \sum_{l=1}^{3}e^{-i\mu_L(n_1,n_2)t}(a_l(n_1,n_2)b_l(n_1,n_2)$$

$$P_{n_1}P_{n_2}-\varepsilon_1 b_l(n_1,n_2)^2\sqrt{n_1+2}P_{n_1+2}P_{n_2}$$

$$-e^{-i\mu_L(n_1+2,n_2)t}\varepsilon_2 a_l(n_1,n_2-2)$$

$$c_l(n_1,n_2-2)\sqrt{n_2}P_{n_1}P_{n_2-2}$$

$$+e^{-i\mu_L(n_1+2,n_2)t}\varepsilon_2\sqrt{n_2+2}$$

$$a_l(n_1+2,n_2)^2 P_{n_1}P_{n_2+2}, \quad (90)$$

$$C_{n_1,n_2,t} =$$
$$\sum_{l=1}^{3}e^{-i\mu_L(n_1,n_2)t}a_l(n_1,n_2)c_l(n_1,n_2)P_{n_1}P_{n_2}$$

$$-\varepsilon_1 b_l(n_1,n_2)c_l(n_1,n_2)\sqrt{n_1+2}P_{n_1+2}P_{n_2})$$

$$+e^{-i\mu_L(n_1,n_2+2)t}\varepsilon_2\sqrt{n_2+2}a_l(n_1,n_2+2)^2 P_{n_1}. \quad (91)$$